\documentclass[12pt]{article}
\usepackage{a4wide,epsfig,amsmath,amssymb,cite,scalefnt} 

\newcommand{\msbar}{$\overline{\rm MS}$}

\newcommand{\qsla}{q\!\!\!/\,\,}
\newcommand{\Qsla}{Q\!\!\!\!/\,\,\,}

\newcommand{\ep}{\varepsilon}
\newcommand{\order}[1]{\mathcal{O}\!\left( #1 \right)}

\newcommand{\lnzone}{L_x}
\newcommand{\lnztwo}{L_x^2}
\newcommand{\lnzthree}{L_x^3}

\newcommand{\zz}{x}

\allowdisplaybreaks[1]

\begin{document}

\title{\vskip-3cm{\baselineskip14pt
    \begin{flushleft}
      \normalsize SFB/CPP-07-47 \\
      \normalsize TTP07-21
  \end{flushleft}}
  \vskip1.5cm
  Light quark mass effects in the on-shell renormalization constants
}
\author{\small S. Bekavac$^{(a)}$, A. Grozin$^{(b)}$, 
  D. Seidel$^{(a)}$ and M. Steinhauser$^{(a)}$\\[1em]
{\small\it (a) Institut f{\"u}r Theoretische Teilchenphysik,
  Universit{\"a}t Karlsruhe (TH)},\\
{\small\it 76128 Karlsruhe, Germany}
\\
{\small\it (b) Budker Institute of Nuclear Physics,}\\
{\small\it Novosibirsk 630090, Russia}
}

\date{}

\maketitle

\thispagestyle{empty}

\begin{abstract}
  We compute the three-loop relation between the pole and the minimally
  subtracted quark mass allowing for virtual effects from
  a second massive quark. We also consider the analogue effects for
  the on-shell wave function renormalization constant.
\medskip


\end{abstract}

\thispagestyle{empty}

\newpage


\section{\label{sec::intro}Introduction}

Quark masses are fundamental parameters of the Standard Model (SM)
and thus it is desirable to determine their numerical values
with the highest possible precision.
In order to do so it is necessary to fix a renormalization scheme
which defines the quark mass.
A renormalization scheme which is of particular importance in those
situations where the quark mass is small as compared to the typical
energy scale is the $\overline{\rm MS}$ scheme. Although it has no 
immediate physical interpretation, in general a good convergence
in perturbation theory is observed.
On the other hand, 
the most intuitive renormalization scheme 
is the on-shell scheme where the
renormalized quark mass is defined as the pole of the propagator.
It is well known that such a definition is plagued by
numerically large long-distance effects which are connected to the 
renormalon corrections.
As a consequence one observes a bad behaviour of the 
perturbative expansion if the pole mass is used as a parameter.
Still, it plays a crucial role in all computations of threshold phenomena 
like energy levels or the decay of 
a quark-anti-quark bound state.
In order to make contact between high-energy phenomena on one side and 
threshold processes on the other side it is necessary to have a
precise relation between the pole and the $\overline{\rm MS}$
quark mass at hand. Important three-loop contributions are provided in
this paper. As a by-product we also compute the same kind of 
corrections to the wave function renormalization constant.

The one-loop corrections to the relation between the 
$\overline{\rm MS}$ and on-shell mass have been computed 
almost 30 years ago in Ref.~\cite{Tarrach:1980up}. 
About ten years later, the two-loop corrections have been
calculated in Ref.~\cite{Gray:1990yh}. Shortly afterwards also the
two-loop result 
for the on-shell wave function renormalization constant $Z_2^{OS}$
has been obtained~\cite{Broadhurst:1991fy}. 
Contributions with a single mass scale
were expressed via master integrals exactly in $d=4-2\varepsilon$; the 
$\ep$-expansion of the only non-trivial master integral is 
known~\cite{Broadhurst:1991fi,Broadhurst:1996az}
far enough to obtain $O(\ep^4)$ terms in $Z_m^{OS}$ and $Z_2^{OS}$.
Contributions with two mass scales were obtained up to $O(1)$ only.
Later, a reduction algorithm for two-loop two-scale on-shell
integrals has been constructed, and expressions via master
integrals exact in $d$ have been 
obtained~\cite{Davydychev:1998si}.\footnote{The $\ep$-expansion of the
  two non-trivial master integrals was 
  only obtained up to $\order{1}$. One of these integrals was later expanded
  to $\order{\ep^5}$~\cite{Argeri:2002wz} (though only the $\order{\ep}$ term
  is published). Both integrals were calculated~\cite{Onishchenko:2002ri} up
  to $\order{\ep}$, as series in the mass ratio up to the sixth order.}
The three-loop relation between the $\overline{\rm MS}$ and on-shell
mass has been computed at the end of the nineties in a semi-numerical way 
in Refs.~\cite{Chetyrkin:1999ys,Chetyrkin:1999qi} where
the off-shell fermion propagator has been considered for small and
large external momenta. The on-shell quantities have been 
obtained with the help of a conformal mapping and Pad\'e
approximation. Later the results have 
been confirmed in Ref.~\cite{Melnikov:2000qh} by an analytical on-shell
calculation. The three-loop result for $Z_2^{\rm OS}$ has been
obtained in Ref.~\cite{Melnikov:2000zc}.
Both analytical calculations have recently been rederived in
Ref.~\cite{Marquard:2007uj}.
In this paper we complete the three-loop results for the
mass counterterm $Z_m^{\rm OS}$ and $Z_2^{\rm OS}$ by computing 
the contributions where
a second mass scale is present through a closed quark loop.
We would like to mention that the approximation linear in the mass
ratio has been derived in Ref.~\cite{Hoang:2000fm} using the
corresponding corrections to the static potential~\cite{Melles:2000dq}
and their cancellation in the relation between the $\overline{\rm MS}$
and 1S quark mass.

We concentrate on the situation where the second quark mass 
is smaller than the external one, although
our formulae can also be applied to the reversed situation.
Then, however, often it is advantageous to perform a decoupling of the heavy
mass leading to an effective theory where the latter is integrated out.

As far as the quark masses in the SM are concerned the corrections
considered in this paper
are of practical relevance for the bottom quark
where the second mass scale is given by the charm quark.
In this case we have for the mass ratio $m_c/m_b\approx 0.3$. Thus the
massless approximation does not provide a good result.
In all other cases the effect from a lighter quark mass is negligible.
Nevertheless, below generic results are presented.

In principle at three-loop order there is also a contribution
involving two additional masses which are present in two 
closed fermion loops. However, for all practical applications 
the lightest quark mass can safely be neglected.

The remainder of the paper is organized as follows: In
Section~\ref{sec::osren} we discuss the framework which is used for
the calculation. Afterwards the results for the
relation between the on-shell and $\overline{\rm MS}$ quark mass
and the wave function renormalization constant are discussed 
in Sections~\ref{sec::zm} and~\ref{sec::z2}, respectively.
Since the analytical expressions are quite involved we present in both
cases handy approximation formulae of our results.
Finally, Section~\ref{sec::concl} contains a simple application and
our conclusions. Appendix~\ref{app::MI}
lists all master integrals in graphical from.


\section{\label{sec::osren}On-shell renormalization of quark mass and wave
  function} 

The formulae relevant for the computation of the renormalization
constants for the mass and wave function have been derived in
Refs.~\cite{Melnikov:2000zc,Marquard:2007uj}. For completeness we
repeat the resulting expressions which read
\begin{eqnarray}
  Z_m^{\rm OS} &=& 1 + \Sigma_1(M_q^2,M_q)\,,
  \label{eq::defs::calcZm} \\
  \left( Z_2^{\rm OS} \right)^{-1} &=& 1 + 2M_q^2
  \frac{\partial}{\partial q^2} \Sigma_1(q^2,M_q) \Big|_{q^2 = M_q^2} +
  \Sigma_2(M_q^2,M_q) \,,
  \label{eq::defs::calcZ2}
\end{eqnarray}
where $Z_m^{\rm OS}$ and $Z_2^{\rm OS}$ are defined through
\begin{eqnarray}
  m_{q,0} &=& Z_m^{\rm OS}\, M_q\,,
  \label{eq::defs::defZm} \\
  \psi_0 &=& \sqrt{Z_2^{\rm OS}}\, \psi\,.
  \label{eq::defs::defZ2}
\end{eqnarray}
$\psi$ is the quark field renormalized in the on-shell scheme
with mass $m_q$, $M_q$ is the 
on-shell mass and bare quantities are denoted by a subscript 0.
$\Sigma$ denotes the quark self-energy
contributions which can be decomposed as
\begin{eqnarray}
  \Sigma(q,m_q) &=&
  m_q\, \Sigma_1(q^2,m_q) + (\qsla - m_q)\, \Sigma_2(q^2,m_q)\,.
  \label{eq::defs::sigmadecomp}
\end{eqnarray}
For completeness let us also introduce the $\overline{\rm MS}$
renormalization constant via
\begin{eqnarray}
  m_{q,0} &=& Z_m^{\overline{\rm MS}}\, m_q(\mu)\,.
  \label{eq::defs::defZmMS}
\end{eqnarray}

The quantities on the right-hand side of Eqs.~(\ref{eq::defs::calcZm})
and~(\ref{eq::defs::calcZ2}) are obtained by considering 
the external momentum of the quarks to be $q = Q(1+t)$
with $Q^2 = M_q^2$. The application of the projector 
$(\Qsla + M_q)/(4M_q^2)$ and an expansion to first order in $t$
leads to
\begin{eqnarray}
  {\rm Tr} \left\{ \frac{\Qsla + M_q}{4M_q^2} \Sigma(q,M_q) \right\} &=&
  \Sigma_1(q^2,M_q) + t \Sigma_2(q^2,M_q) \nonumber \\
  &=& \Sigma_1(M_q^2,M_q)
  + \left( 2M_q^2 \frac{\partial}{\partial q^2} \Sigma_1(q^2,M_q)
  \Big|_{q^2 = M_q^2} \!\!+\! \Sigma_2(M_q^2,M_q) \right) t \nonumber\\ 
  &&  + \order{t^2} \,.
  \label{eq::defs::trace}
\end{eqnarray}
Thus, to obtain $Z_m^{\rm OS}$ one only needs to calculate 
$\Sigma_1$ for $q^2 = M_q^2$. 
To calculate $Z_2^{\rm OS}$, one has to compute the first derivative of
the self-energy diagrams. The mass renormalization is taken into account
iteratively by calculating one- and two-loop diagrams with zero-momentum
insertions.

\begin{figure}[t]
  \leavevmode
  \epsfig{figure=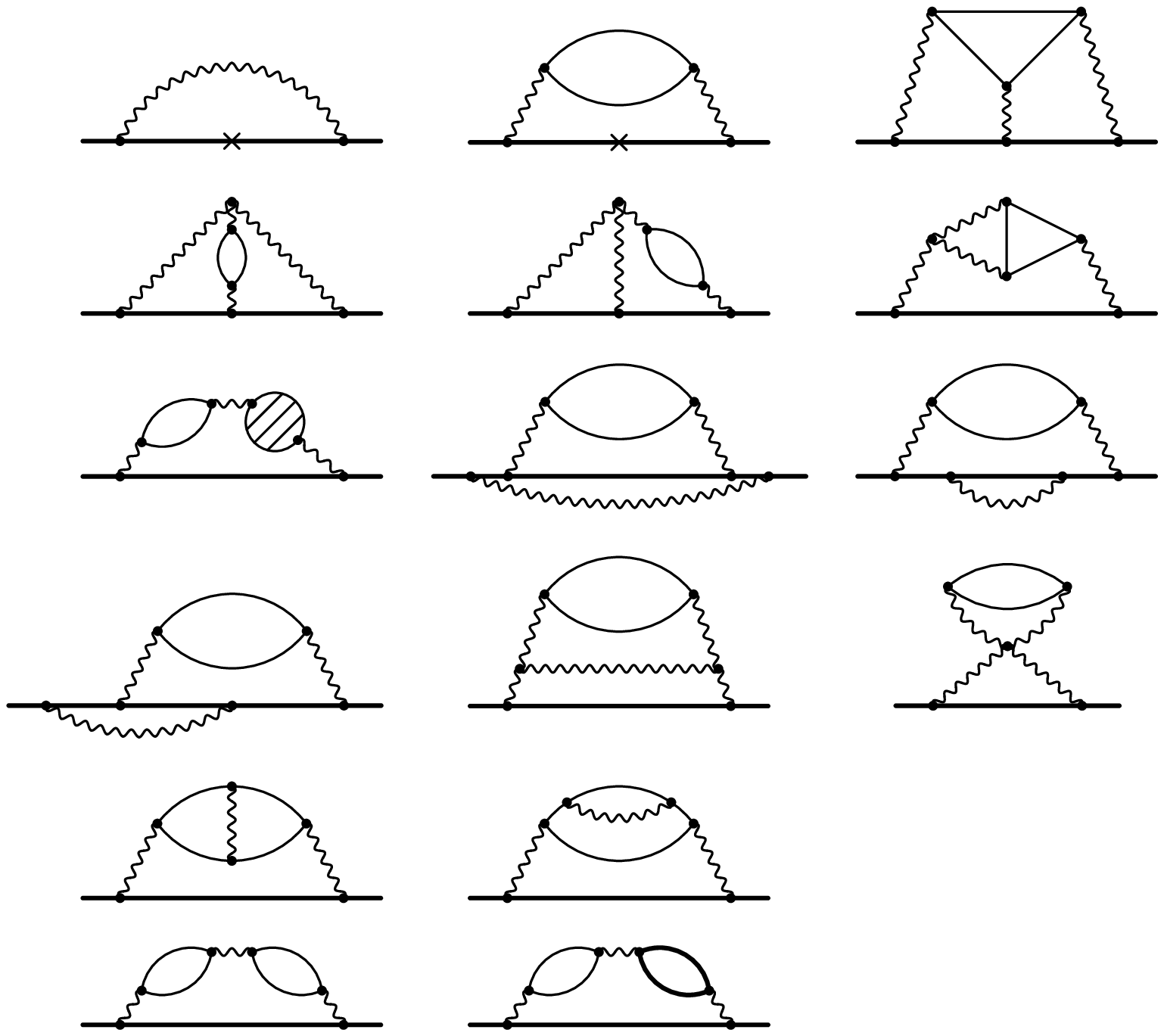,width=\textwidth}
  \caption{\label{fig::diagrams} 
    All three-loop heavy-quark self-energy diagrams containing $n_m$.
    Thick solid lines denote massive quarks with mass 
    $m_q$ and thin ones quarks with mass $m_f$. Wavy lines denote
    gluons and crosses mark counterterm insertions. The shaded blob
    denotes the sum of one-loop massless self-energy 
    insertions (massless quarks, gluons, ghosts).}
\end{figure}

The results for the renormalization constants can be cast into the
following form
\begin{eqnarray}
  Z_i^{\rm OS} &=& 1 + \frac{\alpha_s(\mu)}{\pi} \left(\frac{e^{\gamma_E}}{4 \pi}
  \right)^{-\varepsilon} \delta Z_i^{(1)} +
  \left(\frac{\alpha_s(\mu)}{\pi}\right)^2 \left(\frac{e^{\gamma_E}}{4 \pi}
  \right)^{-2\varepsilon} \delta Z_i^{(2)} \nonumber \\
  && + \left(\frac{\alpha_s(\mu)}{\pi}\right)^3
  \left(\frac{e^{\gamma_E}}{4\pi} \right)^{-3\varepsilon} \delta Z_i^{(3)}
  + \mathcal{O}\left(\alpha_s^4\right) \,,
  \label{eq::mass::Zm}
\end{eqnarray}
with $i\in\{m,2\}$.
It is convenient to further decompose the two- and three-loop contribution in
terms of the different colour factors
\begin{eqnarray}
  \delta Z_i^{(2)}(\zz) &=& C_F^2\, Z_i^{FF} + C_FC_A\, Z_i^{FA} +
  + C_FT_Fn_l Z_i^{FL} + C_FT_Fn_h Z_i^{FH} + C_FT_Fn_m Z_i^{FM}(\zz) 
  \nonumber\\
  \delta Z_i^{(3)}(\zz) &=& C_F^3\, Z_i^{FFF} + C_F^2C_A\, Z_i^{FFA} +
  C_FC_A^2\, Z_i^{FAA} 
  + C_FT_Fn_l \left( C_F\, Z_i^{FFL} + C_A\, Z_i^{FAL}
  \right.
  \nonumber\\&&\mbox{} 
  \left.
  + T_Fn_l\,Z_i^{FLL} + T_Fn_h\, Z_i^{FHL} + T_Fn_m\, Z_i^{FML}(\zz)
  \right) 
  \nonumber\\&&\mbox{} 
  + C_FT_Fn_h \left( C_F\, Z_i^{FFH} + C_A\, Z_i^{FAH} 
  + T_Fn_h\,Z_i^{FHH} + T_Fn_m\,Z_i^{FMH}(\zz) \right) \nonumber\\
  &&\mbox{} + C_FT_Fn_m \left( C_F\, Z_i^{FFM}(\zz) + C_A\, Z_i^{FAM}(\zz) 
  + T_Fn_m\,Z_i^{FMM}(\zz) \right)
  \,,
  \label{eq::mass::Zm3l}
\end{eqnarray}
where $n_l$ and $n_h$ mark the closed quark loops with mass
zero and $m_q$, respectively. $n_m$ labels the closed quark loops
involving a second mass scale which we denote as $m_f$. 
Although we have $n_h=1$ and $n_m=1$ 
in our applications, we keep a generic label which is useful 
when tracing the origin of the individual contributions.
In Eq.~(\ref{eq::mass::Zm3l}) 
$C_F = (N_c^2 - 1)/(2N_c)$ and $C_A = N_c$ are the eigenvalues of
the quadratic Casimir operators of the fundamental and adjoint
representation of $SU(N_c)$, respectively. In the case of QCD we have
$N_c=3$. $T_F = 1/2$ is the index of the fundamental representation and
$n_f=n_l+n_h+n_m$ is the number of quark flavours.
$\alpha_s(\mu)$ is the strong coupling constant defined in the \msbar\
scheme with $n_f$ active flavours.
The coefficients proportional to $n_m$ 
are functions of the mass ratio which we
define by
\begin{eqnarray}
  \zz &=& \frac{M_f}{M_q}
  \label{eq::z}
  \,,
\end{eqnarray}
i.e. the ratio of the on-shell masses. 
The dependence of the coefficients in
Eq.~(\ref{eq::mass::Zm3l}) on $L_\mu=\log\mu^2/M_q^2$ is suppressed. 
The Feynman diagrams producing $n_m$-dependent contributions to $Z_m^{\rm OS}$
and $Z_2^{\rm OS}$ are shown in Fig.~\ref{fig::diagrams}.

The one-loop result for $Z_m^{\rm OS}$
expanded up to order $\varepsilon^2$ can be found in Eq.~(13)
of Ref.~\cite{Marquard:2007uj} and the expression for $\delta
Z_m^{(2)}|_{n_m=0}$ 
including ${\cal O}(\varepsilon)$ terms is given in Eq.~(14) of the same
reference. $Z_m^{FM}$ and $Z_2^{FM}$ have been computed in
analytic form in Ref.~\cite{Gray:1990yh}.
The main result of this paper are the functions
$Z_m^{FFM}$, $Z_m^{FAM}$, $Z_m^{FLM}$, $Z_m^{FHM}$ and $Z_m^{FMM}$
which are discussed in Section~\ref{sec::zm}.

In the case of the mass renormalization it is convenient to
consider the ratio between the on-shell and $\overline{\rm MS}$
renormalization constants
\begin{eqnarray}
  z_m &=& \frac{Z_m^{\rm OS}}{Z_m^{\overline{\rm MS}}} \,\,=\,\,  
  \frac{m_q}{M_q}
\end{eqnarray}
which is finite. We furthermore adopt the notation introduced in 
Eqs.~(\ref{eq::mass::Zm}) and~(\ref{eq::mass::Zm3l})
for $Z_m^{\rm OS}$ also for $z_m$.
Let us for later reference provide already here the
result for the $\overline{\rm MS}$ renormalization constant which is
given by~\cite{Tarasov:1982gk}
\begin{eqnarray}
  Z^{\overline{\rm MS}}_m &=& 1 
+\sum \limits_{i=1}^{\infty} C_i \left ( \frac {\alpha_s(\mu)}{\pi} \right )^i,
\end{eqnarray}
with
\begin{eqnarray}
C_1 &=& -\frac {1}{\varepsilon}, \hspace{5em}
\nonumber\\
C_2 &=& \frac {1}{\varepsilon^2}
         \left ( \frac {15}{8} - \frac {1}{12} n_f \right )
+ \frac {1}{\varepsilon}\left (  - \frac {101}{48} + \frac {5}{72}n_f \right ),
\nonumber \\
C_3 &=&
 \frac {1}{\varepsilon^3}\left (  - \frac {65}{16} + \frac {7}{18}n_f 
 - \frac {1}{108}n_f^2 \right )
 + \frac {1}{\varepsilon^2}\left ( \frac {2329}{288} - \frac {25}{36}n_f 
 + \frac {5}{648}n_f^2 \right )
\nonumber \\
&& + \frac {1}{\varepsilon}
      \left (  - \frac {1249}{192} + \frac {5}{18}\zeta_3 n_f 
   + \frac {277}{648}n_f + \frac {35}{3888}n_f^2 \right ).
\end{eqnarray}

As far as the wave function renormalization constant is concerned, we
have at the one-loop level $\delta Z_2^{(1)}=\delta Z_m^{(1)}$ and the
two-loop contributions for $n_m=0$, including order $\varepsilon$ terms, 
can be found in Eq.~(25) of Ref.~\cite{Marquard:2007uj}.
The results for the functions
$Z_2^{FFM}$, $Z_2^{FAM}$, $Z_2^{FLM}$, $Z_2^{FHM}$ and $Z_2^{FMM}$
are discussed in Section~\ref{sec::z2}.

Starting from the three-loop level, the wave function renormalization
constant depends on the gauge parameter, $\xi$, which we define
through the gluon propagator as
\begin{equation}
  D_{\mu\nu}^{ab}(k) = -\frac{i}{k^2}\, \left( g_{\mu\nu} - \xi\,
  \frac{k_\mu k_\nu}{k^2} \right)\, \delta^{ab}\,,
  \label{eq::wave::gluon}
\end{equation}
where $a$ and $b$ are colour indices.


\section{\label{sec::zm}On-shell mass relation}


\subsection{Two-loop result}

Before discussing in detail the three-loop result let us
consider the two-loop quantity $z_m^{FM}$. The analytical result can
be found in Ref.~\cite{Gray:1990yh} and is given by
\begin{eqnarray}
  z_m^{FM} &=& \frac{1}{96} 
  \Big\{48 \zz^4 \log^2(\zz) +48 \zz^2 \log (\zz) 
  +72 \zz^2+4 L_\mu (3 L_\mu+13)
  \nonumber\\&&\mbox{}
  +8 \pi ^2 \left(\zz^4-3 \zz^3-3 \zz+1\right)+71
  \nonumber\\&&
  \mbox{}
  -48 (\zz+1)^2 \left(\zz^2-\zz+1\right)
  \left[\log (\zz) \log (\zz+1)+\text{Li}_2(-\zz)\right]
  \nonumber\\&&\mbox{}
  -48 (\zz-1)^2 \left(\zz^2+\zz+1\right) \left[\log (\zz)\log (1-\zz) 
  +\text{Li}_2(\zz)\right]
  \Big\}
  \,,
  \label{eq::delm2}
\end{eqnarray}
where $L_\mu=\log(\mu^2/M_q^2)$ and Li$_2$ is the dilogarithm.
In our approach all occuring integrals are reduced to four master
integrals~\cite{Davydychev:1998si}. 
The Harmonic Polylogarithms (HPL)~\cite{Remiddi:1999ew}
which appear in a first step in the results of these
integrals can be transformed into the (di)logarithms of
Eq.~(\ref{eq::delm2}) resulting in complete agreement with
Ref.~\cite{Gray:1990yh}.\footnote{A trivial overall factor $C_F=4/3$ is
  actually missing in Eqs.~(17) and~(20) of this reference.}

\begin{figure}[t]
  \leavevmode
  \epsfxsize=\textwidth
   \epsffile{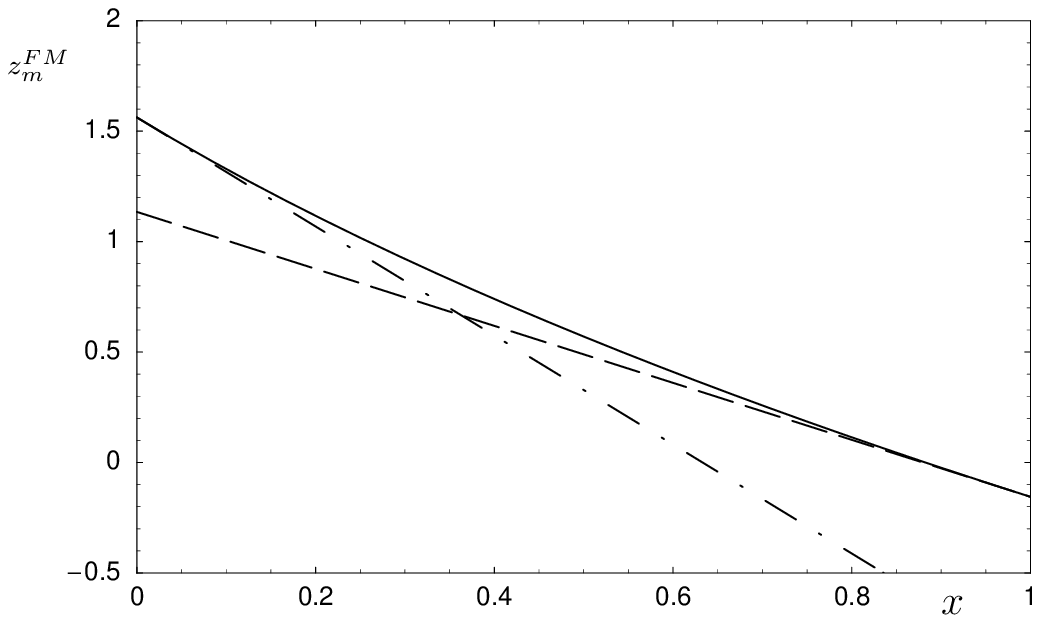}
  \caption{\label{fig::zmFM}The two-loop correction $z_m^{FM}$ as a
    function of $\zz$ at $\mu=M_q$. Next to the exact result (solid
    line) also the  
    approximations for $\zz\to0$  (dash-dotted line) and 
    for $\zz\to1$ (dashed line) are shown including the linear terms.
    }
\end{figure}

In Fig.~\ref{fig::zmFM} $z_m^{FM}$ is shown as a function of $\zz$ for
$\mu^2=M_q^2$. In addition to the exact result we also 
show the curves corresponding to the linear approximations for $\zz\to0$ and 
$\zz\to1$. 
One observes a rapid convergence of both expansions which 
almost extends to $\zz=1$ ($\zz=0$) once the $\zz^{10}$ ($(1-\zz)^{10}$) terms are included.
Note that the $\zz^{5}$ terms are 
sufficient in order to provide an
excellent approximation far below the per mill level for $\zz=0.3$.
We also want to mention that a 12\%
deviation is observed (for $x=0.3$) 
if only the linear terms in $\zz$ are included
into the expansion.

Let us finally provide an approximation
formula~\cite{Gray:1990yh} which agrees to better than 1\% with the
analytical formula of Eq.~(\ref{eq::delm2}):
\begin{eqnarray}
  \tilde{z}_m^{FM} &=& 1.562-2.394\zz+0.9428\zz^2-0.2666\zz^3
   +\frac{13}{24}L_\mu+\frac{1}{8}L_\mu^2
  \,.
\end{eqnarray}


\subsection{Three-loop corrections}

At three-loop order we want to discuss in a first step the
coefficients of the five $n_m$-dependent colour structures but afterwards
also consider the physical quantity which is obtained after inserting
the numerical values for the QCD colour factors.

All Feynman diagrams are generated with {\tt
QGRAF}~\cite{Nogueira:1991ex} and the various topologies are identified 
with the help of {\tt q2e} and {\tt
exp}~\cite{Harlander:1997zb,Seidensticker:1999bb}.
In a second calculation the three-loop diagrams were generated
starting from three generic topologies
which already appear in the calculation of Ref.~\cite{Grozin:2006xm}.
In a next step the reduction of the various functions to so-called master
integrals (MI's) has to be achieved. For this step we use the so-called
Laporta method~\cite{Laporta:1996mq,Laporta:2001dd} 
which reduces the three-loop integrals to 27
MI's. We use the implementation of Laporta's
algorithm in the program {\tt Crusher}~\cite{PMDS}. 
It is written in {\tt C++} and uses
{\tt GiNaC}~\cite{Bauer:2000cp} for simple
manipulations like taking derivatives of polynomial quantities. In the
practical implementation of the Laporta algorithm one of the most
time-consuming operations is the simplification of the coefficients
appearing in front of the individual integrals. This task is performed
with the help of {\tt Fermat}~\cite{fermat} where a special interface
has been used (see Ref.~\cite{Tentyukov:2006ys}).
The main features of the
implementation are the automated generation of the
integration-by-parts (IBP) identities~\cite{Chetyrkin:1981qh}, 
a complete symmetrization of the diagrams and the possibility to use multiprocessor
environments.

In Figs.~\ref{fig::master1}--\ref{fig::master4} of Appendix~\ref{app::MI}
a graphical representation of the master
integrals can be found.
We have chosen two independent ways to compute the
$\varepsilon$-expansion of the master integrals. The first one relies
on the Mellin-Barnes technique (see, e.g., Ref.~\cite{Smirnov:2004ym})
and provides us with numerical results. Here we have used the {\tt Mathematica}
package {\tt MB.m}~\cite{Czakon:2005rk}. With the help of
our second method, based on differential equations, Ref.~\cite{Kotikov:1990kg},
we were able to evaluate all but four
master integrals in analytic form.
More details can be found in Ref.~\cite{BGSS}.

The coefficient functions of all master integrals
contain a $1/\varepsilon$ pole, some even a $1/\varepsilon^2$ term which
means that the master integrals have to be expanded up to order
$\varepsilon$ and $\varepsilon^2$, respectively.
All but four master integrals could be evaluated analytically in terms
of HPLs  which we
evaluate numerically with the help of the {\tt Mathematica} package {\tt
  HPL.m}~\cite{Maitre:2005uu,Maitre:2007kp}. 
Two of the remaining four integrals, which are
all needed up to order $\varepsilon$, could be computed including the 
constant term~\cite{BGSS}
and for the residual two integrals analytical results
are obtained for the pole parts. For the still remaining six coefficients
integral representations are available which in the worst case are
two-dimensional. 

Close to $\zz=0$ we observe large cancellations between the contributions
originating from different master integrals. 
On the other hand, as we have seen above, the expansion for $\zz\ll1$
converges very fast; at two-loop order the first five terms 
approximate the exact result to 0.02\% for $\zz=0.3$ which is relevant
for the charm mass effects to the bottom quark mass.
For this reason we decided to derive an expansion of our result 
including terms of order $\zz^{8}$. The results expressed in terms of
HPLs can simply be expanded using {\tt
  HPL.m}~\cite{Maitre:2005uu,Maitre:2007kp}. For the remaining 
coefficients we use their Mellin-Barnes representation in order to
express them in terms of multiple sums which in turn leads to the
coefficients of $\zz^n$.

\begin{figure}[t]
  \leavevmode
  \epsfxsize=\textwidth
  \epsffile{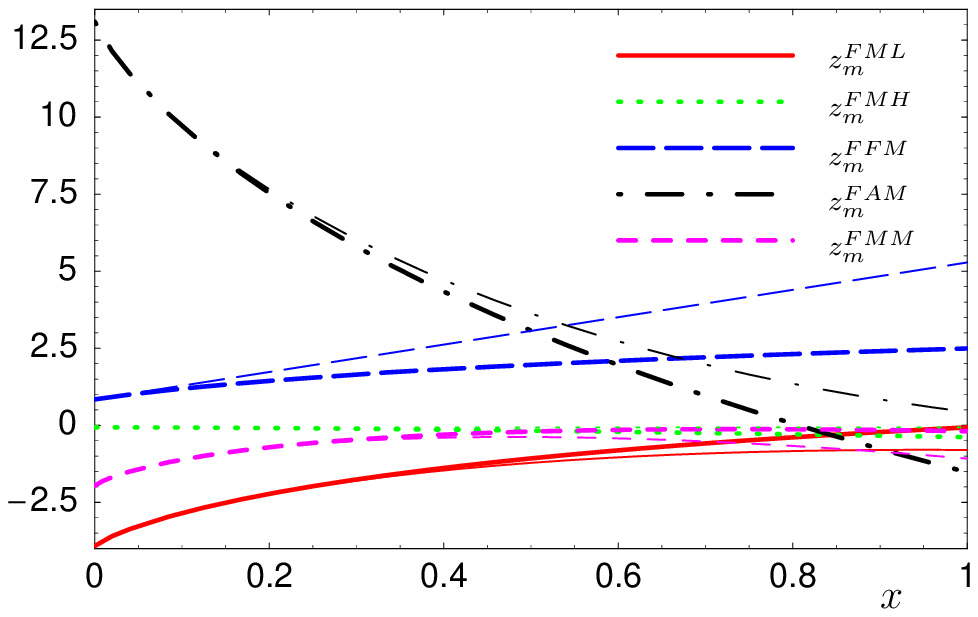}
  \caption{\label{fig::zm3l}
    $z_m^{FFM}$, $z_m^{FAM}$, $z_m^{FLM}$, $z_m^{FHM}$ and $z_m^{FMM}$
    as a function of $\zz$.
    The exact curves are represented by the thick and the linear
    approximations for small $\zz$ by the thin lines.
  }
\end{figure}

Our result for the five functions
$z_m^{FFM}$, $z_m^{FAM}$, $z_m^{FLM}$, $z_m^{FHM}$ and $z_m^{FMM}$
are shown in Fig.~\ref{fig::zm3l} for $0\le \zz\le1$.
The exact results are represented by thick lines and the 
expansion terms for $\zz\to0$ up to the linear term as thin lines.
The latter provide a good approximation to
the exact results up to about 
$\zz\approx0.1 \ldots 0.4 $, depending on the colour structure.
We want to mention that the expansion terms including
corrections of order $\zz^8$ provide a good approximation almost up
to $\zz=1$.
Note that for $\zz=0.3$ the linear approximation deviates from the
exact result (obtained by the proper sum of the individual colour structures) 
by 8\%, whereas the deviation including terms up to $\zz^5$ is only
0.008\%.

The $\zz$-dependence of the individual curves is rather flat, 
in particular close to $\zz=1$ --- except for $z_m^{FAM}$. The latter
varies from 
$+13.10$ for $\zz=0$ to $-1.52$ for $\zz=1$ with 
a zero for $\zz\approx 0.81$.
It is interesting to note that the coefficient of $C_F^2Tn_m$ is
positive for $0\le \zz\le1$ whereas the contributions originating from
the diagrams involving two closed fermion loops are negative
and furthermore numerically smaller in a large part of the $\zz$ interval.

The boundary terms for $\zz=0$ and $\zz=1$ can be extracted from the 
known single-scale three-loop 
results~\cite{Chetyrkin:1999qi,Chetyrkin:1999ys,Melnikov:2000qh,Marquard:2007uj}
and are given by
\begin{align}
  z_m^{FFM}(0) &= z_m^{FFL} \approx 0.842  \,,
  & z_m^{FFM}(1) &= z_m^{FFH} \approx 2.494\,,\nonumber\\
  z_m^{FAM}(0) &= z_m^{FAL} \approx 13.099 \,,
  & z_m^{FAM}(1) &= z_m^{FAH} \approx -1.522\,,\nonumber\\
  z_m^{FLM}(0) &= 2z_m^{FLL} \approx -3.916 \,,
  & z_m^{FLM}(1) &= z_m^{FHL} \approx -0.067\,,\nonumber\\
  z_m^{FHM}(0) &= z_m^{FHL} \approx  -0.067 \,,
  & z_m^{FHM}(1) &= 2z_m^{FHH} \approx -0.384\,,\nonumber\\
  z_m^{FMM}(0) &= z_m^{FLL} \approx -1.958  \,,
  & z_m^{FMM}(1) &= z_m^{FHH} \approx  -0.192  \,.
\end{align}
These limits constitute an important cross check of our calculation.
We find perfect agreement with the known results.

We refrain from listing explicit results for the $\zz$-dependent
coefficients which can be obtained at the URL
{\tt http://www-ttp.physik.uni-karlsruhe.de/Progdata/ttp07/ttp\\07-21/}.
Instead we provide approximation formulae which have an accuracy of
better than 1\% for $0\le \zz\le1$. 
They are inspired by the expansion
for small values of $\zz$ and the behaviour for 
$\zz=1$.\footnote{We want to mention that the formulae provided on the web
  page are valid for $0\le \zz\le5$.}
We obtain
\begin{eqnarray}
 \tilde{z}_m^{FFM} &=& 0.842+4.333\zz-1.365\zz^2+3.136\zz^2\log 
\zz-1.316\zz^3\nonumber\\
   &&+L_\mu\left(-0.972+1.820\zz-0.811\zz^2+0.279\zz^3\right)
     -\frac{13}{32}L_\mu^2-\frac{3}{32}L_\mu^3\,,\nonumber\\
 \tilde{z}_m^{FAM} &=& 13.099-12.945\zz+9.041\zz\log \zz-1.676\zz^2 
\nonumber\\
   &&+L_\mu\left(6.724-4.407\zz+1.807\zz^2-0.549\zz^3\right)
     +\frac{373}{288}L_\mu^2+\frac{11}{72}L_\mu^3\,,\nonumber\\
 \tilde{z}_m^{FLM} &=& -3.916+2.948\zz-3.304\zz\log \zz+0.901\zz^2 
\nonumber\\
   &&+L_\mu\left(-1.921+1.604\zz-0.660\zz^2+0.201\zz^3\right)
     -\frac{13}{36}L_\mu^2-\frac{1}{18}L_\mu^3\,,\nonumber\\
 \tilde{z}_m^{FHM} &=& -0.067-0.612\zz^2+0.443\zz^3-0.148\zz^4 \nonumber\\
   &&+L_\mu\left(-0.776+1.597\zz-0.631\zz^2+0.179\zz^3\right)
     -\frac{13}{36}L_\mu^2-\frac{1}{18}L_\mu^3\,,\nonumber\\
 \tilde{z}_m^{FMM} &=& -1.958+0.501\zz-3.403\zz\log 
\zz+1.264\zz^2-0.103\zz^3\log \zz \nonumber\\
   &&+L_\mu\left(-0.960+1.600\zz-0.653\zz^2+0.198\zz^3\right)
     -\frac{13}{72}L_\mu^2-\frac{1}{36}L_\mu^3\,.
  \label{eq::zm-zexp}
\end{eqnarray}

\begin{figure}[t]
  \leavevmode
  \epsfxsize=\textwidth
  \epsffile{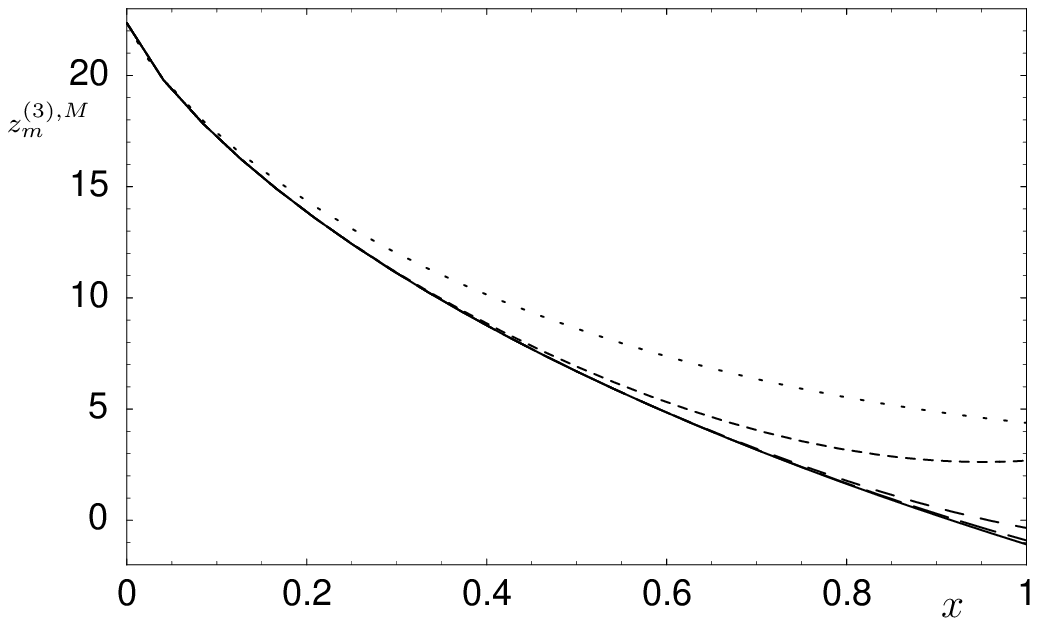}
  \caption{\label{fig::zmM}
    $z_m^{(3),M}$
    as a function of $\zz$ where $n_l=3$ and $\mu^2=M_q^2$ has been chosen. 
    The exact result is is shown together with the 
    the expansions
    up to $\zz^n$ ($n=1,3,5,8$) (dotted line to long-dashed line).
  }
\end{figure}

Let us in the following consider the result in the case of QCD, i.e.,
we set $C_F=4/3, C_A=3, T_F=1/2, n_h=1, n_m=1$ and define the quantities
\begin{eqnarray}
  \label{eq::mass::zM}
  z_m^{(2),M} &=& C_F T_F z_m^{FM}
  \,,
  \label{eq::zm3M}
  \\
  z_m^{(3),M} &=&
  C_F^2T_F   \, z_m^{FFM} 
  + C_FC_AT_F\, z_m^{FAM}
  + C_FT_F^2 \, z_m^{FML}
  + C_FT_F^2 \, z_m^{FMH}
  + C_FT_F^2 \, z_m^{FMM} 
  \,.\nonumber
\end{eqnarray}
Evaluating the coefficients for the $z$-expansion in numerical form
the results become very compact and are given by ($\mu^2=M_q^2$)
\begin{eqnarray}
  z_m^{(2),M} &=&\mbox{}
+\zz^{0}\left(   1.0414\right)
+\zz^{1}\left(  -1.6449\right)
+\zz^{2}\left(   1.0000\right)
+\zz^{3}\left(  -1.6449\right)
\nonumber\\&&\mbox{}
+\zz^{4}\left(   1.2474 -    0.7222 \lnzone +    0.3333 \lnztwo\right)
+\zz^{6}\left(  -0.0844 +    0.0889 \lnzone\right)
\nonumber\\&&\mbox{}
+\zz^{8}\left(  -0.0118 +    0.0214 \lnzone\right)
\,,
\nonumber\\
  z_m^{(3),M} &=&\mbox{}
+\zz^{0}\left(  26.2712 -    1.3054 n_l\right)
\nonumber\\&&\mbox{}
+\zz^{1}\left( -21.0921 +   16.9977 \lnzone +    1.0385 n_l -    1.0966 \lnzone n_l\right)
\nonumber\\&&\mbox{}
+\zz^{2}\left(  12.7021 +   10.6870 \lnzone -    0.2222 n_l\right)
\nonumber\\&&\mbox{}
+\zz^{3}\left( -13.0084 +   16.5103 \lnzone -    0.2408 n_l -    1.0966 \lnzone n_l\right)
\nonumber\\&&\mbox{}
+\zz^{4}\left[  -4.1035 +    0.1938 \lnzone - 0.2593 \lnzthree +0.7919 \lnztwo
+ n_l\left(0.3613 - 0.2822 \lnzone 
\right.\right.\nonumber\\&&\left.\left.\qquad\mbox{}
+ 0.0741 \lnzthree - 0.2407 \lnztwo \right)\right] 
\nonumber\\&&\mbox{}
+\zz^{5}\left(  -1.4908 +    2.8512 \lnzone +    0.4935 n_l\right)
\nonumber\\&&\mbox{}
+\zz^{6}\left(   0.1654 -    0.5756 \lnzone +    0.8224 \lnztwo -    0.1873 n_l +    0.0267 \lnzone n_l\right)
\nonumber\\&&\mbox{}
+\zz^{7}\left(   -0.1751 +    0.5452 \lnzone  +    0.0653 n_l\right)
\nonumber\\&&\mbox{}
+\zz^{8}\left[  -0.0705   +   0.0492 \lnzone  +    0.3125 \lnztwo 
  +n_l\left(-  0.0377 +  0.0025 \lnzone \right)\right]
  \,,
  \label{eq::zmM}
\end{eqnarray}
where the contributions proportional to $n_l$ are listed separately
and $\lnzone=\log \zz$.
In Fig.~\ref{fig::zmM} the expansions of the quantity $z_m^{(3),M}$
up to $\zz^n$ ($n=1,3,5,8$) are shown 
together with the exact expressions where $n_l=3$ has been chosen
corresponding to the case $M_f=M_c$ and $M_q=M_b$.
One observes a rapid convergence when including successively higher orders.
In Tab.~\ref{tab::zm} numerical results
for the individual coefficient functions, for $z_m^{(3),M}$ and
$z_m^{(3)}$ are shown in the region around $\zz=0.3$ where again $n_l=3$
has been adopted.

\begin{table}[t]
\begin{center}
\begin{tabular}{r|rrrrr|r|r}
  $\zz$ & $z_m^{FFM}$ & $z_m^{FAM}$ & $z_m^{FLM}$ & $z_m^{FHM}$ & $z_m^{FMM}$ &
  $z_m^{(3)M}$ & $z_m^{(3)}$
  \\
  \hline
$   0.26 $&$   1.57 $&$   6.45 $&$  -1.93 $&$  -0.10 $&$  -0.55 $&$  12.15 $&$ -111.66 $ \\
$   0.27 $&$   1.59 $&$   6.29 $&$  -1.89 $&$  -0.10 $&$  -0.52 $&$  11.89 $&$ -111.92 $ \\
$   0.28 $&$   1.61 $&$   6.12 $&$  -1.84 $&$  -0.11 $&$  -0.50 $&$  11.62 $&$ -112.19 $ \\
$   0.29 $&$   1.63 $&$   5.96 $&$  -1.80 $&$  -0.11 $&$  -0.48 $&$  11.37 $&$ -112.44 $ \\
$   0.30 $&$   1.64 $&$   5.80 $&$  -1.76 $&$  -0.11 $&$  -0.46 $&$  11.11 $&$ -112.70 $ \\
$   0.31 $&$   1.66 $&$   5.64 $&$  -1.72 $&$  -0.11 $&$  -0.44 $&$  10.86 $&$ -112.95 $ \\
$   0.32 $&$   1.68 $&$   5.49 $&$  -1.68 $&$  -0.12 $&$  -0.42 $&$  10.61 $&$ -113.20 $ \\
$   0.33 $&$   1.70 $&$   5.34 $&$  -1.64 $&$  -0.12 $&$  -0.41 $&$  10.37 $&$ -113.44 $ \\
$   0.34 $&$   1.72 $&$   5.19 $&$  -1.60 $&$  -0.12 $&$  -0.39 $&$  10.13 $&$ -113.68 $ \\
\end{tabular}
\caption{\label{tab::zm}
  Numerical results for $\zz$-dependent coefficients contributing
  to the three-loop quantity $z_m^{(3)}$.
  For the renormalization scale $\mu^2=M_q^2$ has been adopted.}
\end{center}
\end{table}

Quite often it is convenient to consider the masses which only appear
in closed loops in the $\overline{\rm MS}$ scheme. Thus, transforming 
$M_f$ to the $\overline{\rm MS}$ scheme leads to the following 
modifications of Eq.~(\ref{eq::mass::zM}):
\begin{eqnarray}
  z_m^{(2),M}(\zz) &\to& z_m^{(2),M}(\zz_f)
  \,,\nonumber \\
  z_m^{(3),M}(\zz) &\to& z_m^{(3),M}(\zz_f) + C_F^2 T_F n_m
  \Delta z_m^f(\zz_f)
  \,,
\end{eqnarray}
with
\begin{eqnarray}
  \zz_f &\equiv& \zz_f(\mu_f) \,\,=\,\, \frac{m_f(\mu_f)}{M_q}
  \,,\nonumber \\
  \Delta z_m^f(\zz_f) &=& \frac{1}{24} 
  \left(\frac{3}{2} l_{\mu_f} +2 \right) 
  \zz \big\{24 \zz^3\log ^2(\zz) 
  +12 \zz\log (\zz) +24 \zz
  \nonumber\\&&\mbox{}
  +\pi^2 \left[\zz^2 (4 \zz-9)-3\right]
  -6 \left[(4 \zz+3) \zz^2+1\right] \left[\log (\zz) \log
  (\zz+1)+\text{Li}_2(-\zz)\right]
  \nonumber\\&&\mbox{}
  -6 (\zz-1) \left(4 \zz^2+\zz+1\right) \left[\log (1-\zz) \log (\zz)
  +\text{Li}_2(\zz)\right]\big\}
  \,,
  \label{eq::zbar}
\end{eqnarray}
where $l_{\mu_f} = \log(\mu_f^2/m_f^2)$.
In Eq.~(\ref{eq::zbar}) we introduced the scale $\mu_f$ for the
renormalization point of the quark mass $m_f$ which is
different from $\mu$ implicitly present in Eq.~(\ref{eq::mass::Zm3l}).
The latter contains the information about the running of $\alpha_s$
whereas the former incorporates the anomalous mass dimension of $m_f$.

Let us for completeness also present the inverse relation $1/z_m$ which
is conveniently expressed in terms of the $\overline{\rm MS}$ quark
mass $m_q$. Using analogous conventions to~(\ref{eq::mass::zM}) we find
\begin{eqnarray}
  \label{eq::zmq}
  (1/z_m)^{(2),M}(\zz_q) &=& 
  (M_q/m_q(\mu))^{(2),M}(x_q) = -
  z_m^{(2),M}(\zz_q)\Big|_{L_\mu\to l_\mu}
  \,,\nonumber \\
  (1/z_m)^{(3),M}(\zz_q) &=& 
  (M_q/m_q(\mu))^{(3),M}(x_q) = -
  z_m^{(3),M}(\zz_q)\Big|_{L_\mu\to l_\mu} 
  + C_F^2 T_F n_m \Delta z_m^q(\zz_q)
  \,,\nonumber \\
\end{eqnarray}
with
\begin{eqnarray}
  \zz_q &\equiv& \zz_q(\mu) \,\,=\,\, \frac{M_f}{m_q(\mu)}
  \,,\nonumber \\
  \Delta z_m^q(\zz_q) &=& \frac{1}{192} 
  \big\{48 (3 l_\mu+7) \zz_q^4 \log ^2(\zz_q) 
  +144 \zz_q^2 \log (\zz_q) +312 \zz_q^2
  \nonumber\\&&\mbox{}
  +8 \pi ^2 \left(7 \zz_q^4-15 \zz_q^3-3 \zz_q
  -1\right)+137
  \nonumber\\ &&\mbox{}
  -l_\mu \left[-72 \zz_q^2+24 l_\mu \left(\frac{3}{2}l_\mu+4\right)
  +12 \pi ^2 \left(-2 \zz_q^4+3 \zz_q^3-3 \zz_q+2\right)+13\right]
  \nonumber\\ &&\mbox{}
  -48 \left[(7 \zz_q+5) \zz_q^3+\zz_q+\frac{3}{2} l_\mu 
  \left(2 \zz_q^4+\zz_q^3-\zz_q-2\right)-1\right]
  \nonumber\\ &&\mbox{}
  \hspace{2em}\times\left[\log (\zz_q) \log (\zz_q+1)
  +\text{Li}_2(-\zz_q)\right]
  \nonumber\\ &&\mbox{}
  +48 (\zz_q-1) \left[\frac{3}{2} l_\mu 
  \left(2 \zz_q^3+\zz_q^2+\zz_q+2\right)
  +7\zz_q^3 + 2 \zz_q^2 + 2 \zz_q + 1
  \right]
  \nonumber\\ &&\mbox{}
  \hspace{2em}\times\left[-\log (1-\zz_q)
  \log(\zz_q)-\text{Li}_2(\zz_q)\right]
  \big\} 
  \,.
  \label{eq::zq}
\end{eqnarray}
It is understood that the renormalization scale dependent logarithms
appearing in $z_m^{(2),M}(\zz_q)$, $z_m^{(3),M}(\zz_q)$ and 
$\Delta z_m^q(\zz_q)$ are defined as $l_\mu=\log(\mu^2/m_q^2)$.

Again we can consider the masses which only appear
in closed loops in the $\overline{\rm MS}$ scheme, which leads to the following 
modifications of Eq.~(\ref{eq::zmq}):
\begin{eqnarray}
  (1/z_m)^{(2),M}(\zz_q) &\to& (1/z_m)^{(2),M}(\zz_{fq})
  \,,\nonumber \\
  (1/z_m)^{(3),M}(\zz_q) &\to& (1/z_m)^{(3),M}(\zz_{fq})
  + C_F^2 T_F n_m 
  \Delta z_m^{fq}(\zz_{fq})
  \,,
\end{eqnarray}
with
\begin{eqnarray}
  \zz_{fq} &\equiv& \zz_{fq}(\mu_f,\mu) \,\,=\,\, \frac{m_f(\mu_f)}{m_q(\mu)}
  \,,\nonumber \\
  \Delta z_m^{fq}(\zz_{fq}) &=& \frac{1}{24} 
  \left(\frac{3}{2} l_{\mu_f}+2\right) \zz_{fq} 
  \big\{-24 \zz_{fq}^3\log ^2(\zz_{fq}) -12 \zz_{fq}\log (\zz_{fq}) 
  -24 \zz_{fq}
  \nonumber\\&&\mbox{}
  +\pi ^2 \left[(9-4 \zz_{fq}) \zz_{fq}^2+3\right]
  \nonumber\\&&\mbox{}
  +6 \left[(4 \zz_{fq}+3) \zz_{fq}^2+1\right] \left[\log (\zz_{fq}) \log
  (\zz_{fq}+1)+\text{Li}_2(-\zz_{fq})\right]
  \nonumber\\&&\mbox{}
  +6 (\zz_{fq}-1) \left[4 \zz_{fq}^2+\zz_{fq}+1\right] \left[\log (1-\zz_{fq}) 
  \log (\zz_{fq})+\text{Li}_2(\zz_{fq})\right]\big\}
    \,.
    \label{eq::zfq}
\end{eqnarray}

At the end of this Section we want to compare our result with the one
of Ref.~\cite{Hoang:2000fm} where $z_m^{(3),M}$ has been computed 
in the linear approximation. Our result for the linear term reads
\begin{eqnarray}
  (M_f/m_q(m_q))^{(3),M}(x_q)\big|_{\rm linear} &=& 
  x_q \bigg[19.996-16.998\log x_q
  + n_l\left(-1.039+1.097\log x_q \right)
  \bigg]\,. 
  \nonumber\\
\end{eqnarray}
We find agreement for three of the terms but the coefficient with
the numerical value $19.996$ takes the value $21.277$ 
in~\cite{Hoang:2000fm}.\footnote{Note that $n_l$ as introduced in
  Ref.~\cite{Hoang:2000fm} corresponds to our combination $n_l+n_m$.}
This difference can be explained by the approximations performed in
Ref.~\cite{Hoang:2000fm} in order to extract the linear term of the
mass relation.\footnote{We thank A. Hoang for
  communications on this point.}



\section{\label{sec::z2}Wave function renormalization constant}

In contrast to $Z_m^{\rm OS}$ the wave function renormalization constant
contains next to ultraviolet also infrared divergences. Thus it is not
possible to construct a finite quantity by considering the ratio
between the on-shell and $\overline{\rm MS}$ renormalization constant.
For this reason we discuss in what follows the coefficients 
of the $\varepsilon$-expansion separately.

\subsection{Two-loop result}

The two-loop corrections to $Z_2^{(2),M}$ have been computed in
Ref.~\cite{Broadhurst:1991fy}. We confirmed this result and obtain
\begin{eqnarray}
  Z_2^{(2),M} &=&   \frac{1}{\varepsilon} \left( \frac{1}{24} - \frac{1}{3}
  \log x \right)
  + \frac{1}{4}L_\mu^2+\left(
  \frac{1}{6\varepsilon}+\frac{11}{36}-\frac{2}{3} \log\zz \right)L_\mu
  \nonumber \\ 
  & &\mbox{}+ \frac{443}{432} + \frac{5\,\pi^2}{72}
  -\frac{\pi^2}{4}\,x + \frac{7}{6}\,x^2 -\frac{5\,\pi^2}{12}\,x^3
  +\frac{\pi^2}{6}\,x^4 \nonumber \\ 
  & &\mbox{}+ \left( \frac{4}{9} + \frac{2}{3}\,x^2 \right)\,\log x 
  +\left( \frac{2}{3} + x^4 \right) \log^2 x \nonumber \\
  & &\mbox{}+\left( -\frac{1}{3} + \frac{1}{2}\,x +\frac{5}{6}\,x^3 - x^4
  \right)\,\left[ \log x\,\log(1-x) + \text{Li}_2(x) \right]
  \nonumber \\
  & &\mbox{}-\left( \frac{1}{3} + \frac{1}{2}\,x +\frac{5}{6}\,x^3 + x^4
  \right)\, \left[ \log x\,\log(1+x) + \text{Li}_2(-x) \right]\,.
  \label{eq::delZ2}
\end{eqnarray}
The convergence properties are very similar to $z_m^{(2),M}$ and shall
not be discussed here. However, we would like to present a handy
approximation formula which is obtained by an interpolation
where the logarithmic divergence for $\zz\to0$ is extracted
before. It reads
\begin{eqnarray}
  Z_2^{(2),M} &=& 
  \frac{1}{\varepsilon} \left( \frac{1}{24} - \frac{1}{3} \log x \right) 
  + \frac{1}{4}L_\mu^2+\left(
  \frac{1}{6\varepsilon}+\frac{11}{36}-\frac{2}{3} \log\zz \right)L_\mu
  \nonumber\\&&\mbox{}
  +\frac{2}{3} \log^2 x+\frac{4}{9} \log x 
  +1.711-2.356 x+1.125 x^2 -0.344 x^3
  \,.
  \label{eq::delZ2approx}
\end{eqnarray}
and works to better than 1\%  for $x \in [0,1]$.

\subsection{Three-loop result}

We again refrain from listing explicit results for the $x$-dependent
coefficients and refer to the URL
{\tt http://www-ttp.physik.uni-karlsruhe.de/Progdata/ttp07/ttp07-21/}
where the expressions can be downloaded in {\tt Mathematica} format.
It is, however, useful to present results for the analogue quantity to 
$z_m^{(3),M}$ as defined in Eq.~(\ref{eq::zm3M}).
The cubic and quadratic poles can be presented analytically and read
\begin{eqnarray}
  Z_2^{(3),M}\Big|_{\varepsilon^{-3}} &=& \frac{1-\xi}{96}
  \,,\nonumber\\
  Z_2^{(3),M}\Big|_{\varepsilon^{-2}} &=&
  - \frac{23}{108} - \frac{89}{96} L_\mu
  + \frac{19}{16}\log\zz
  + n_l \left(\frac{1}{108} + \frac{1}{36} L_\mu
  - \frac{1}{18} \log\zz\right)
  \nonumber\\&&\mbox{}
  + \xi\left(\frac{1}{32} - \frac{1}{32} L_\mu
  + \frac{1}{16}\log\zz\right) 
  \,.
\end{eqnarray}
Concerning the single pole and the finite part we again present
handy approximation formulae which we obtain by an interpolation to
our expression after subtracting the singular terms for
$\zz\to 0$.
We cast the result in the form
\begin{eqnarray}
  Z_2^{(3),M}\Big|_{\varepsilon^0} &=&    
  a_0 + a_1 L_\mu + a_2 L_\mu^2 + a_3 L_\mu^3 
  + \xi\, \left( b_0 + b_1 L_\mu + b_2 L_\mu^2 + b_3 L_\mu^3 \right)\,,
  \nonumber \\ 
  Z_2^{(3),M}\Big|_{\varepsilon^{-1}} &=&     
  c_0 + c_1 L_\mu + c_2 L_\mu^2
  + \xi\, \left( d_0 + d_1 L_\mu + d_2 L_\mu^2\right) \,,
\end{eqnarray}
where the coefficients are given by
\begin{eqnarray}
  a_0 &=& 0.38426 \log (x) \left[\log (x)+0.53600\right] 
  \left[\log (x)+13.51219\right]
  \nonumber\\&&\mbox{}  
  +25.383-22.326 x+11.127x\log (x)-1.473 x^2
  \,,\nonumber\\
  a_1 &=&  \log (x)\left[-1.46528 \log (x)-5.54630\right]
  +6.832-1.971x+0.982x^2-0.324x^3
  \,,\nonumber\\
  a_2 &=& \frac{403}{288}\,\log (x)+\frac{127}{48}
  \,,\nonumber\\
  a_3 &=&-\frac{193}{576}
  \,,\nonumber\\
  b_0 &=& \frac{407}{864}+\frac{\pi ^2}{128}-\frac{7 \zeta
    (3)}{96}+\left(\frac{35}{48}+\frac{\pi
      ^2}{64}\right) \log (x)+\frac{9}{16}\, \log^2(x)
  +\frac{3}{8}\, \log ^3(x)
  \,,\nonumber\\
  b_1 &=&-\frac{35}{96}-\frac{\pi
    ^2}{128}-\frac{9}{16}\, \log (x)-\frac{9}{16}\, \log ^2(x)
  \,,\nonumber\\
  b_2 &=& \frac{9}{64}+\frac{9}{32}\, \log (x) 
  \,,\nonumber\\
  b_3 &=&-\frac{3}{64}
  \,,\nonumber\\
  c_0 &=& \log (x)\left[-1.34028 \log (x)-2.41667\right] 
  -1.352+2.367x-1.180x^2+0.387x^3
  \,,\nonumber\\
  c_1 &=&\frac{23}{36}+\frac{257}{144}\, \log (x)
  \,,\nonumber\\
  c_2 &=&-\frac{41}{64}
  \,,\nonumber\\
  d_0 &=&-\frac{35}{288}-\frac{\pi
    ^2}{384}-\frac{3}{16}\, \log (x)-\frac{3}{16}\, \log ^2(x)
  \,,\nonumber\\
  d_1 &=& \frac{3}{32}+\frac{3}{16}\, \log (x)
  \,,\nonumber\\
  d_2 &=&-\frac{3}{64}
  \,.
\end{eqnarray}
For the singular contributions, which we know to high precision, we provide
five digits after the decimal point whereas for the results from the fit three
digits are given.

As in the case of $z_m$ we want to present the expanded
results for $\zz\to 0$ which are given by ($\mu^2=M_q^2$)
\begin{eqnarray}
  Z_2^{(2),M} &=& \frac{1}{\varepsilon}\left(\frac{1}{24}-\frac{1}{3}\,\log
    (x)\right) 
  +\frac{443}{432}+\frac{5 \pi ^2}{72}+\frac{4}{9}\, \log (x)+\frac{2}{3}\,
  \log ^2(x)
  \nonumber\\&&\mbox{} 
  -\frac{\pi ^2}{4}\,x+2 x^2-\frac{5 \pi ^2}{12}\,x^3
  +\left(\frac{125}{72}+\frac{\pi ^2}{6}-\frac{11}{6}\, \log(x)+\log
    ^2(x)\right) x^4
  \nonumber\\&&\mbox{}
  +\left(-\frac{22}{75}+\frac{16}{45}\, \log (x)\right) x^6
  + \left(-\frac{379}{7840}+\frac{3}{28}\, \log (x)\right) x^8
  \,,
  \\
  Z_2^{(3),M} &=& \frac{1}{\varepsilon^3}\bigg[0.0104-0.0104 \xi\bigg]
  \nonumber\\
  &&\mbox{}      + \frac{1}{\varepsilon^2}\bigg[ -0.213+(0.0093-0.0556 L_x) n_l+1.1875 L_x+(0.0625
    L_x+0.0313) \xi\bigg]
  \nonumber\\
  &&\mbox{}      + \frac{1}{\varepsilon}\bigg[ -1.3978
  +\left(0.0151+0.0741 L_x+0.0556L_x^2\right) n_l
  -2.6389 L_x-1.5069 L_x^2
  \nonumber\\
  &&\quad\mbox{}
  +2.4674 x-2x^2+4.1123 x^3+\left(-3.381+1.8333 L_x-L_x^2\right) x^4
  \nonumber\\
  &&\quad\mbox{} +(0.2933-0.3556 L_x) x^6
  +(0.0483-0.1071 L_x) x^8
  \nonumber\\
  &&\quad\mbox{}  +\left(-0.1472-0.1875L_x-0.1875 L_x^2\right) \xi
  \bigg]
  \nonumber\\
  &&\mbox{}  +31.2973+\left(-1.9715+0.1765 L_x-0.0741 L_x^2+0.037 L_x^3\right)
  n_l 
  \nonumber\\
  &&\mbox{}+2.2534 L_x+5.6204L_x^2+0.2731 L_x^3
  \nonumber\\
  &&\mbox{}+\Big[-14.695+19.328 L_x+(1.5578-1.6449 L_x) n_l\Big] x\nonumber\\
  &&\mbox{}+\left(24.2836-0.7778 n_l+27.2951 L_x-0.6389 L_x^2\right)
  x^2\nonumber\\ 
  &&\mbox{}
  +\Big[-23.8364+28.4056 L_x+(-2.7416 L_x-0.6021) n_l\Big] x^3\nonumber\\
  &&\mbox{}+\Big[-12.2965+7.3074 L_x-2.4541L_x^2+1.5787 L_x^3
  +\left(0.527-0.8466L_x
    \right.\nonumber\\&&\left.\quad\mbox{}
    -0.7222 L_x^2+0.2222 L_x^3\right) n_l\Big]
  x^4
  +(1.0836+1.7272 n_l+8.6633 L_x) x^5\nonumber\\
  &&\mbox{}+\Bigg[-2.8058-0.1342L_x+3.125 L_x^2+0.1389 L_x^3+(-0.7078+0.077
    L_x) n_l\Bigg] x^6\nonumber\\ 
  &&\mbox{}+(0.4472+0.2937n_l+1.6073 L_x) x^7
  +\Bigg[-0.9511+0.8787 L_x+1.4563 L_x^2
  \nonumber\\&&\mbox{}
  +0.0995 L_x^3+(-0.1831+0.0054L_x) 
    n_l\Bigg] x^8
  \nonumber\\
  &&\mbox{}+\left(0.4605+0.8834L_x+0.5625 L_x^2+0.375 L_x^3\right) \xi 
  \,.
\end{eqnarray}

\begin{figure}[t]
  \leavevmode
 \begin{center}
  \epsfxsize=\textwidth
   \epsffile{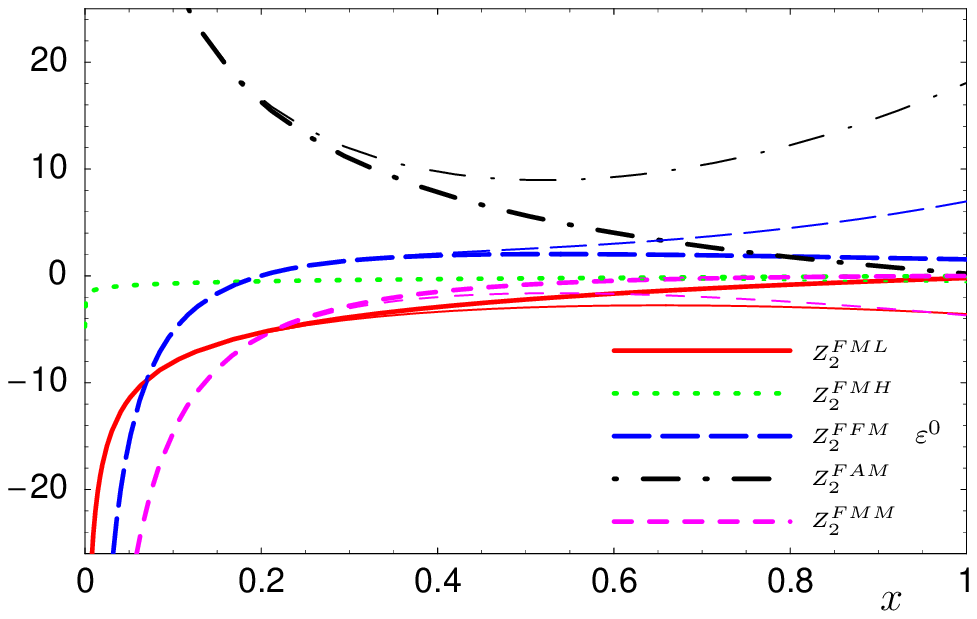}
\end{center}
  \caption{\label{fig::Z23l}
    Finite parts of
    $Z_2^{FML}, Z_2^{FMH}, Z_2^{FFM}, Z_2^{FAM}, Z_2^{FMM}$ 
    for Feynman gauge ($\xi=0$) and $L_\mu=0$
    as a function of $x$.
    The exact curves are represented by the thick and the small-$x$
    approximations by the thin lines.
  }
\end{figure}

\begin{figure}[t]
  \leavevmode
 \begin{center}
  \epsfxsize=\textwidth
  \epsffile{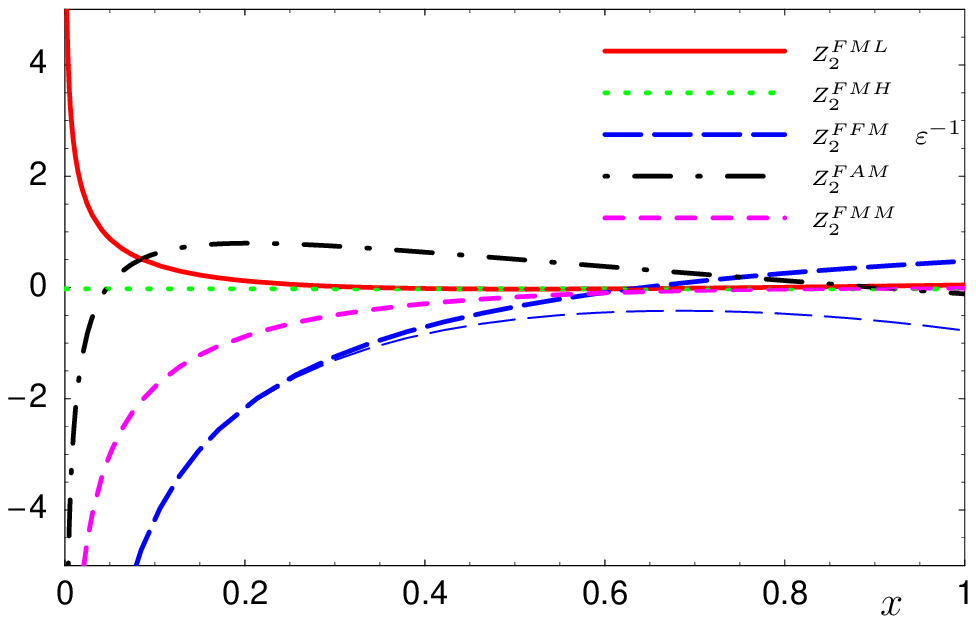}
\end{center}
  \caption{\label{fig::Z23lpole}
    Same as in Fig.~\ref{fig::Z23l} but for the $1/\varepsilon$ pole.
  }
\end{figure}

In Fig.~\ref{fig::Z23l} we compare the exact results for the individual colour
structures (for $\xi=0$) with the approximations
including terms up to order $x^2$. Only for $x\gtrsim0.2$ a difference
is visible.

For completeness we show the analogue curves for the coefficient
of the $1/\varepsilon$ pole in Fig.~\ref{fig::Z23lpole}. It is interesting to
mention that only $Z_2^{FFM}$ has a non-trivial $\zz$-dependence beyond the
logarithmic divergences for $\zz\to0$.


\section{Applications and conclusions}
\label{sec::concl}

As an application of our result we want to discuss the charm quark
effects in the relations between the pole, the $\overline{\rm MS}$ and
the $1S$ quark mass.
For illustration we use $m_b(m_b)=4.2$~GeV, $m_c(m_c)=1.3$~GeV,
$\mu=m_b$ 
and\footnote{As a starting point we use $\alpha_s^{(5)}(M_Z)=0.118$
  and perform the running and decoupling with the program {\tt
    RunDec}~\cite{Chetyrkin:2000yt}. 
  As in Ref.~\cite{Hoang:1998nz} we consider the
  mass relations with four active flavours.}
$\alpha_s^{(4)}(m_b)=0.2247$. The relation between the on-shell
and the $\overline{\rm MS}$ mass leads to
\begin{eqnarray}
  M_b &=& \left[ 4.2 + 0.401 + \left(0.199 + 0.0094\Big|_{m_c}\right)
  + \left(0.145 + 0.0182\Big|_{m_c}\right) \right] \mbox{GeV}
  \,,
  \label{eq::mosmms}
\end{eqnarray}
where the tree-level, one-, two- and three-loop results are shown
separately. The contributions from the charm quark mass which 
vanish for $m_c\to 0$ are marked by a
subscript $m_c$. One observes that the higher order contributions 
are significant. In particular, the two-loop
charm quark effects amount to 9~MeV and the three-loop ones to 
18~MeV. A similar bad convergence is observed in the relation between
the $1S$ mass~\cite{Hoang:1998nz} $M_b^{1S}$ and the pole mass
$M_b$. For $M_b=4.8$~GeV, $m_c(m_c)=1.3$~GeV, $\mu=M_b$ and 
$\alpha_s^{(4)}(M_b)=0.2150$ it is given by
\begin{eqnarray}
  M_b^{1S} &=& \left[4.8 - 0.049 - \left(0.073 + 0.0041\Big|_{m_c}\right)
  - \left(0.098 + 0.0112\Big|_{m_c}\right)\right] \mbox{GeV}
  \,.
  \label{eq::m1smos}
\end{eqnarray}
However, the relation between the $1S$ and the $\overline{\rm MS}$
quark mass is much better behaved as can be seen in the following
example where we have chosen $M_b^{1S}=4.69$~GeV, $m_c(m_c)=1.3$~GeV,
$\mu=M_b^{1S}$ and $\alpha_s^{(4)}(M_b^{1S})=0.2167$
\begin{eqnarray}
  m_b &=& \left[ 4.69 - 0.382 - \left(0.098 + 0.0047\Big|_{m_c}\right)
  - \left(0.030 + 0.0051\Big|_{m_c}\right) \right] \mbox{GeV}
  \,.
  \label{eq::mmsm1s}
\end{eqnarray}
The two-loop charm effects amount to only 4.7~MeV and three-loop ones
to 5.1~MeV. We want to mention that in case only the linear
approximation~\cite{Hoang:2000fm} of the charm quark mass effects is used 
the corresponding three-loop results in Eqs.~(\ref{eq::mosmms}) 
and~(\ref{eq::mmsm1s}) read $0.0167$ and $0.0037$, respectively.

We are now in the position to compare with Eq.~(168) of
Ref.~\cite{Hoang:2000fm} which provides the relation between the $1S$
and $\overline{\rm MS}$ bottom quark mass allowing for a variation of 
$M_b^{1S}$, $m_c$, $\alpha_s$ and the renormalization scale $\mu$.
Updating the coefficient of the $m_c$ term one obtains the formula
\begin{equation}
m_b = 4.169~\mbox{GeV} - 0.009 \left( m_c(m_c) - 1.4~\text{GeV} \right)
  \label{eq::mmsm1s-gen}
\end{equation}
which has an accuracy of better than 0.01\% for
$1.1~\text{GeV}<m_c<1.7~\text{GeV}$. 

A further application of our result would be the incorporation of our
corrections in the analysis of the bottom quark mass determination
from the $\Upsilon(1S)$ system. In the analysis performed in
Ref.~\cite{Penin:2002zv} the charm quark mass has not been considered
and an uncertainty of $\pm10$~MeV has been assigned which could be
reduced to a large extend.

To conclude,
in this paper we have computed the three-loop QCD corrections to the
on-shell renormalization constants for a heavy quark mass and the
corresponding wave function where a second massive quark appears in
closed loops. The two-scale three-loop diagrams are analytically
reduced to 27 master integrals. The $\varepsilon$-expansion of the latter
is computed in analytical form, except for six coefficients for
which one- and two-dimensional integral representations are available.
We derived a compact expansion of the renormalization constants in the limit
where the second quark mass is small. A rapid convergence is observed
providing a good approximation to the exact result even close to 
the equal-mass case.


\bigskip
\noindent
{\large\bf Acknowledgements}\\[.3em]
We thank V.A. Smirnov for cross checks on some of the intermediate
results.
This work was supported by the DFG through SFB/TR~9 and the Graduiertenkolleg
``Hochenergiephysik und Teilchenastrophysik''.


\begin{appendix}

\newpage

\section{\label{app::MI}Master integrals}

In this Appendix we collect the master integrals appearing in our
calculation in graphical form.

\begin{figure}[ht]
  \begin{center}
  \leavevmode
  \epsfxsize=.8\textwidth
  \epsffile[100 670 500 750]{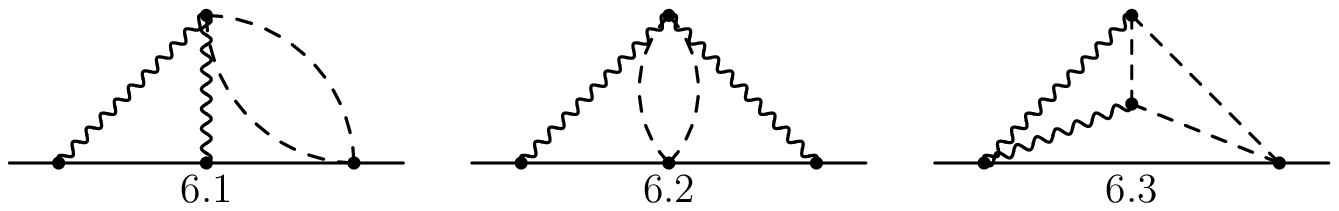}
  \caption{\label{fig::master1} Three-loop master integrals with six
    lines. Solid and dashed
    lines denote massive lines with masses $m_q$ and $m_f$,
    respectively. Wavy lines are massless scalar propagators.}
  \end{center}
\end{figure}

\begin{figure}[ht]
  \begin{center}
  \leavevmode
  \epsfxsize=.6\textwidth
  \epsffile[150 470 420 740]{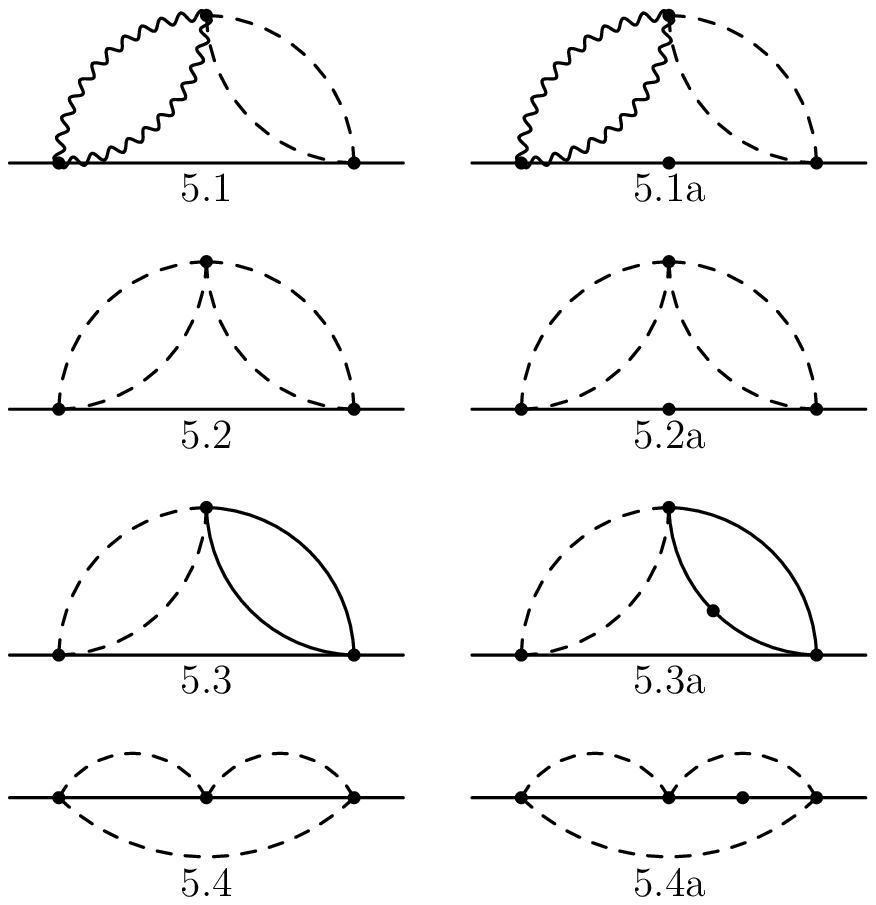}
  \caption{\label{fig::master2} Three-loop master integrals with five
    lines.}
  \end{center}
\end{figure}

\begin{figure}[ht]
  \begin{center}
  \leavevmode
  \epsfxsize=.8\textwidth
  \epsffile[100 350 490 740]{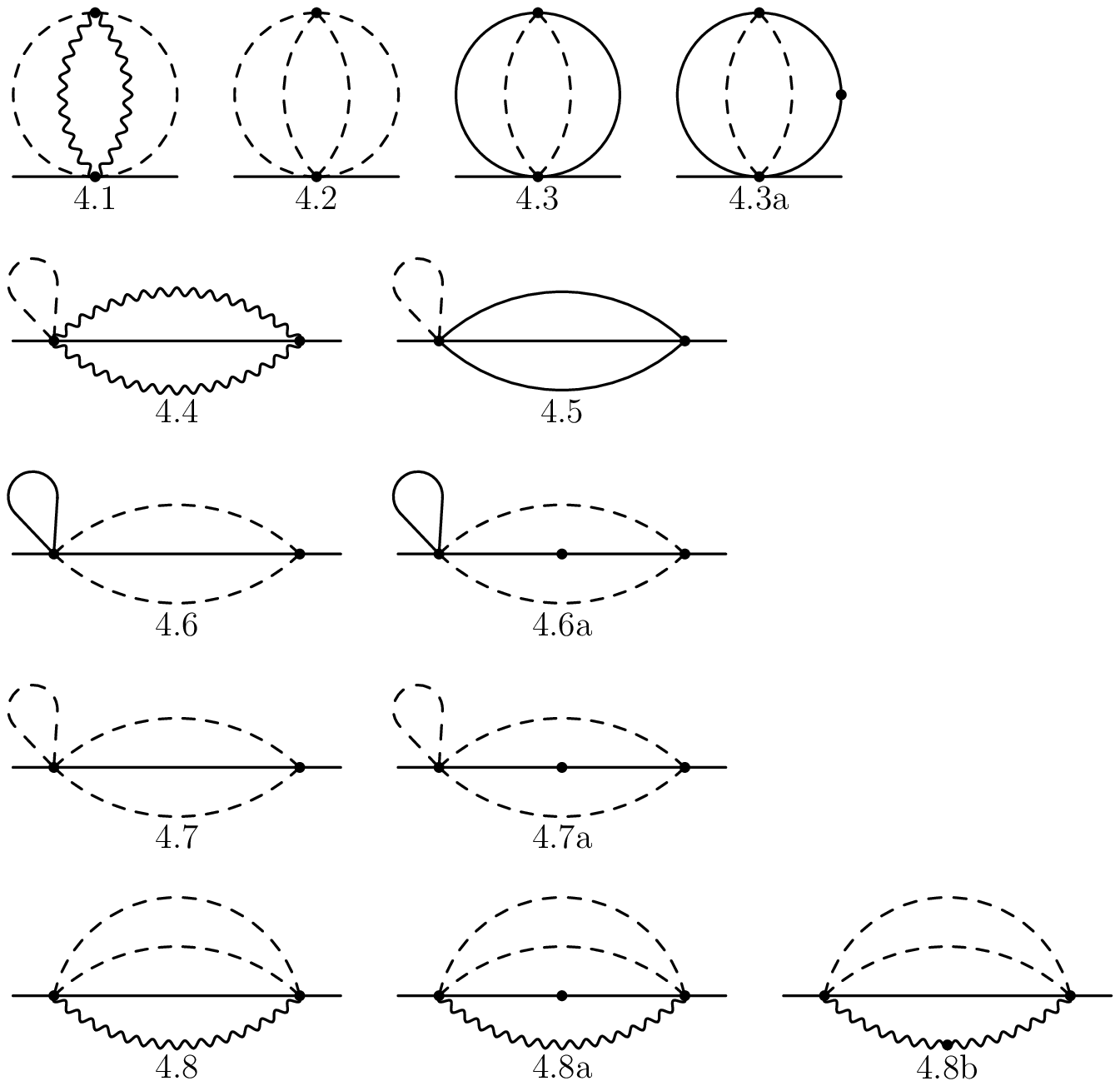}
  \caption{\label{fig::master3} Three-loop master integrals with four
    lines.}
  \end{center}
\end{figure}

\begin{figure}[ht]
  \begin{center}
  \leavevmode
  \epsfxsize=.8\textwidth
  \epsffile[100 630 480 730]{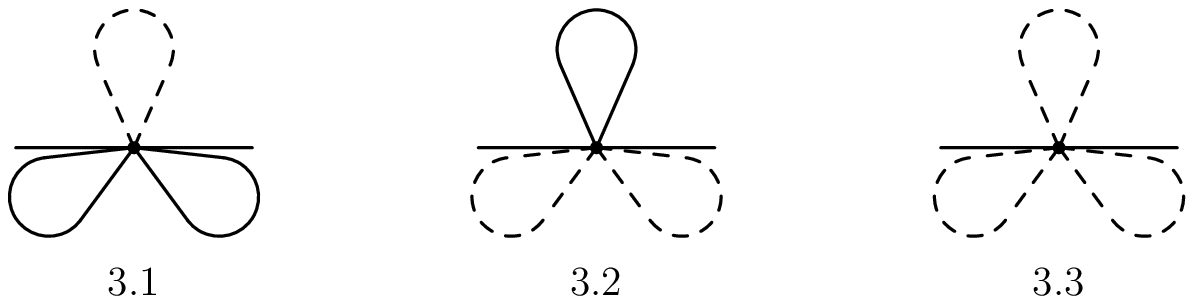}
  \caption{\label{fig::master4} Three-loop master integrals with three
    lines.}
  \end{center}
\end{figure}


\end{appendix}


\newpage



\end{document}